\documentclass[allclo,secnumeq,appnumeq]{FBSart}
\usepackage{amsfonts}
\usepackage{amssymb}
\def\mib#1{\hbox{\boldmath $#1$}}
\def\mbf#1{\hbox{\boldmath $#1$}}

\def\bp{{\mbf p}}
\def\bq{{\mbf q}}
\def\bk{{\mbf k}}

\def\SM3{\Sigma N (3/2)}
\def\SN1{\Sigma N (1/2)}

\def\TS1{\hbox{}^3S_1}
\def\TD1{\hbox{}^3D_1}

\def\eq#1{Eq.~(\ref{#1})}

\def\VRGM{V^{\rm RGM}(\varepsilon)}
\def\VOCM{V^{\rm OCM}(\varepsilon)}
\def\VRGMA{V^{\rm RGM}_\alpha(\varepsilon_\alpha)}
\def\VRGMB{V^{\rm RGM}_\beta(\varepsilon_\beta)}
\def\VRGMC{V^{\rm RGM}_\gamma(\varepsilon_\gamma)}

\def\TTE{\widetilde{T}^{(3)}_\alpha(E, \varepsilon_\alpha)}

\def\erf{{\rm erf}}

\title{Solving Three-Cluster OCM Equations in the Faddeev Formalism}
\author{Y. Fujiwara\instnr{1},
M. Kohno\instnr{2} and
Y. Suzuki\instnr{3}} 
\instlist{Department of Physics, Kyoto University, 
Kyoto 606-8502, Japan
\and Physics Division, Kyushu Dental College,
Kitakyushu 803-8580, Japan
\and Department of Physics, Niigata University,
Niigata 950-2181, Japan
}

\runningauthor{Y.\,Fujiwara, M.\,Kohno
and Y.\,Suzuki}
\runningtitle{Solving Three-Cluster OCM Equations in the Faddeev Formalism}
\begin{document}

\maketitle
\begin{abstract}
Two different types of orthogonality condition models (OCM)
are equivalently formulated in the Faddeev formalism.
One is the OCM which uses pairwise orthogonality conditions
for the relative motion of clusters, and the other is the one
which uses the orthogonalizing pseudo-potential method.
By constructing a redundancy-free $T$-matrix, one can exactly
eliminate the redundant components of the total wave function
for the harmonic-oscillator Pauli-forbidden states,
without introducing any limiting procedure. As an example,
a three-$\alpha$-particle model interacting
via the deep $\alpha \alpha$ potential
by Buck, Friedrich and Wheatley is investigated.
\end{abstract}

\section{Introduction}

The Faddeev formalism for three composite particles has always
suffered from the insufficient treatment of the Pauli principle.
For example, in the three-alpha ($3\alpha$) Faddeev study
by one of the authors \cite{FU67}
one-term separable $\alpha \alpha$ potentials are used to reproduce
the damped inner oscillations of the relative wave functions,
which are the most important effect of the Pauli principle
between two $\alpha$ ($2\alpha$) clusters.
Due to the insufficient treatment of the Pauli principle
among $3\alpha$ clusters, some of the obtained states (${0_2}^+$ and
$1^{-}$ states) were concluded to be spurious since they contain
a large amount of redundant components. On the other hand, a large
binding energy of the shell-model like ground state
and the excited $0^+$ state with well-developed cluster structure
are simultaneously reproduced,
which can never be realized by Ali-Bodmer's phenomenological
$\alpha \alpha$ potential with the repulsive core.
The large overbinding of the $3\alpha$ ground state
with $E_{3\alpha}=-17$ MeV (without the Coulomb force) is most
easily understood by considering that the damped inner oscillations
of the $2\alpha$ relative motion are enhanced
in the compact $3\alpha$ system, and the attractive nature
in the short-range part of the core-less potential
overwhelms the large kinetic energies.\,\cite{SP80}
The same situation is also found in the work by Oryu
and Kamada \cite{OKFB}.
They started directly from the microscopic $2\alpha$-cluster kernel 
of the resonating-group method (RGM), and converted it
to the many-rank separable potentials
which are suitable for the Faddeev calculations.
Although their admixture of the redundant components is relatively
small, the ground-state energy is extremely large
with $E_{3\alpha} \sim -20$ MeV. \cite{KO86}
Since the fish-bone optical model proposed by
Schmid \cite{SC80} also gives a large ground-state energy,
these authors claim that some sort of $3\alpha$ force
is definitely necessary in the $3\alpha$-particle model
to obtain a reasonable agreement
with the experimental observation.\,\cite{OK89}
We think that this repulsive 3-cluster force can be partly avoided
by the complete elimination of the $3\alpha$ redundant components,
which cannot be excluded at the $2\alpha$ potential level.
From the microscopic viewpoint based on the $3\alpha$ RGM,
the model space for the relative motion of the $3\alpha$ clusters
has a well-defined notion solely determined from the assumed
internal wave functions of the $\alpha$ clusters.
The standard interpretation of the 3-cluster force may be
the one which stems from the interaction kernel connected
to the full antisymmetrization among the three clusters.\,\cite{MU95}

Recently, we have developed 3-cluster Faddeev formalism
which employs 2-cluster RGM kernel directly.\,\cite{TRGM,RED}
In this formulation, we first write down the RGM equation
in the form of the Schr{\"o}dinger-type equation.
The resultant interaction term becomes non-local
and energy-dependent.
This linear energy dependence in the interaction term
originates from splitting the overlap kernel
into the direct term and the exchange normalization kernel.
The two-cluster RGM equation sometimes involves
redundant components. In such a case,
the complete off-shell $T$-matrix
is not well-defined in the standard procedure.\,\cite{GRGM}
Our strategy is to distinguish between the energy $\varepsilon$ in
the interaction term and the starting energy $\omega$ involved
in the 2-cluster Green function.
Assuming the energy $\varepsilon$ involved in the interaction term
as a mere parameter, we can define
the full $T$-matrix, $T(\omega,\varepsilon)$, through the  
standard procedure. Although $T(\omega,\varepsilon)$ is singular
when $\omega=\varepsilon$, there is no harm in solving the Faddeev
equation for the bound states since $\omega$ is negative
and $\varepsilon$ is usually positive.
Our finding is that the modified $T$-matrix,
$\widetilde{T}(\omega,\varepsilon)$, obtained by subtracting this
divergent term is the proper ``RGM $T$-matrix'', which should be
used in the Faddeev equation. The remaining energy-dependence
in $\widetilde{T}(\omega,\varepsilon)$ should be
determined self-consistently
by calculating the expectation value of the 2-cluster Hamiltonian
with the resultant total wave-function of the Faddeev equation.
In Ref.\,\cite{TRGM}, we proved that this formalism is
completely equivalent to the 3-cluster orthogonality condition
model (OCM) with pairwise orthogonality conditions (as
proposed by Horiuchi \cite{HO74,HO75}),
using the RGM interaction term as the pairwise interaction.
The energy dependence in the interaction term should be determined 
self-consistently even in this 3-cluster equation.

In this paper, we will show that this Faddeev formalism developed
for 2-cluster RGM kernel is also applicable
to the usual 3-cluster OCM.
In this case, the corresponding ``OCM $T$-matrix'' no longer
involves the $\varepsilon$-dependence, although the
interaction term of the Schr{\"o}dinger-type OCM equation is
still energy dependent. We will also show
that this 3-cluster Faddeev equation
is equivalent to the Faddeev equation formulated
for the ``redundancy-free'' $T$-matrix obtained
from the original OCM potential
by applying Kukulin's method of orthogonalizing
pseudo-potentials \cite{KU78}.
Through these procedures, we can prove the equivalence
between Horiuchi-type 3-cluster OCM with the pairwise orthogonality
conditions and the method of orthogonalizing pseudo-potentials.
A nice point of the present approach is that one no longer needs
the process to take the strength parameter,
$\lambda$ in \eq{form30}, infinity,
which may cause a serious numerical instability
if the model space to solve the 3-cluster equation is too small
in the variational-type calculations.
We think that this is a great merit of using the $T$-matrix
formalism. As a typical example, we investigate the $3\alpha$ system
in which the pair $\alpha$'s interact via the deep local
potential proposed by Buck, Friedrich
and Wheatley (BFW potential) \cite{BF77}.
According to the general idea of the point-like $\alpha$-particle
models, the redundant components of the $2\alpha$ system
are assumed to be the bound states of the $2\alpha$ Hamiltonian.
In this case, the Pauli projection operator $P$ needs a careful
treatment to select a physical model space.
We find that the ground-state energy of this system
is far below the experimental value.
This is a different conclusion from that
reached in Refs.\,\cite{TB03} and \cite{DE03}, which was based
on the variational calculations in the method of orthogonalizing
pseudo-potentials.

In the next section, the Faddeev formalism for the 3-cluster OCM is
developed after a brief recapitulation
of the previous Faddeev formalism
using the 2-cluster RGM kernel \cite{TRGM,RED}.
An application to the BFW potential with the bound-state
Pauli-forbidden states is also given, together with a new feature
influenced by the almost forbidden components of the Faddeev equation.
The relationship of the $T$-matrices derived
in this particular case is easily generalized
for more general types of the Pauli-forbidden states
composed of the harmonic-oscillator (h.o.) wave functions.
The third section discusses the numerical examples of the present
Faddeev formalism for the $3\alpha$ system interacting
via the BFW potential. The last section is devoted to a summary.
A simple formula for $T$-matrices is given in the Appendix.

\section{Formulation}

\subsection{3-cluster Faddeev equation
using the 2-cluster RGM kernel}

We start from a two-cluster RGM equation for the relative
wave function $\chi$, expressed as
\begin{eqnarray}
\left[\,\omega-H_0-V^{\rm RGM}(\omega)\,\right] \chi=0\ ,
\label{form1}
\end{eqnarray}
where $\omega$ is the total energy
in the center-of-mass (c.m.) system,
measured from the two-cluster threshold, $\omega=E-E^{\rm int}$,
$H_0$ is the relative kinetic-energy operator, and
\begin{eqnarray}
V^{\rm RGM}(\omega)=V_{\rm D}+G+\omega K\ ,
\label{form2}
\end{eqnarray}
is the RGM kernel composed of the direct
potential $V_{\rm D}$, the sum
of the exchange kinetic-energy and interaction kernels,
$G=G^{\rm K}+G^{\rm V}$,
and the exchange normalization kernel $K$. For simplicity,
we assume a single-channel RGM and that there exists
only one Pauli-forbidden state $|u\rangle$ (normalized
as $\langle u| u \rangle=1$), which satisfies the
eigen-value equation $K |u\rangle=\gamma |u\rangle$ with
the eigen-value $\gamma=1$.
The projection operator on the Pauli-allowed space for the relative
motion is denoted by $\Lambda=1-|u\rangle \langle u|$.
Using the basic property of the Pauli-forbidden state $|u\rangle$,
$(H_0+V_{\rm D}+G)|u\rangle=0$ and $\langle u|(H_0+V_{\rm D}+G)=0$,
we find that \eq{form1} is equivalent to 
\begin{equation}
\Lambda \left[ \omega-H_0-V^{\rm RGM}(\omega)\right]\Lambda \chi=0
\label{form3}
\end{equation}
or
\begin{equation}
\Lambda \left[ \omega-H_0-v(\omega)\right]\Lambda \chi=0\ \ .
\label{form4}
\end{equation}
Here, we have defined
\begin{equation}
v(\varepsilon)=\Lambda V^{\rm RGM}(\varepsilon) \Lambda
=\Lambda (V_{\rm D}+G) \Lambda+\varepsilon \Lambda K \Lambda\ \ ,
\label{form5}
\end{equation}
by generalizing $\omega$ in $v(\omega)$ to $\varepsilon$.
The OCM approximation consists of
\begin{equation}
\Lambda \left[ \omega-H_0-V_{\rm D}\right]\Lambda \psi=0\ \ ,
\label{form6}
\end{equation}
or more favorably changing the direct potential $V_{\rm D}$ to a 
suitable effective local potential $V_{\rm eff}$.\,\cite{SA68,SA77}

The basic procedure to define the ``RGM $T$-matrix'' is
to separate $\VRGM$ into two distinct parts
\begin{equation}
\VRGM=V(\varepsilon)+v(\varepsilon)
\label{form7}
\end{equation}
with
\begin{equation}
V(\varepsilon)=(\varepsilon-H_0)-\Lambda (\varepsilon-H_0) \Lambda
=\varepsilon |u\rangle \langle u| +\Lambda H_0 \Lambda-H_0\ \ ,
\label{form8}
\end{equation}
and to assume $\varepsilon$ in $v(\varepsilon)$ as
a mere parameter which should be determined
by the surroundings of the interacting two clusters.  
Then we can derive a formal solution of the $T$-matrix equation
\begin{equation}
T(\omega,\varepsilon)=\VRGM+\VRGM\,G^{(+)}_0(\omega)
\,T(\omega,\varepsilon)
\label{form9}
\end{equation}
with $G^{(+)}_0(\omega)=1/(\omega-H_0+i0)$ as follows: \cite{TRGM}
\begin{eqnarray}
T(\omega,\varepsilon) & = & \widetilde{T}(\omega, \varepsilon)
+(\omega-H_0)|u\rangle {1 \over \omega-\varepsilon}
\langle u|(\omega-H_0) \ ,\nonumber \\
\widetilde{T}(\omega, \varepsilon) & = & T_v(\omega, \varepsilon)
-\left[\,1+T_v(\omega, \varepsilon)
G^{(+)}_0(\omega)\,\right] |u \rangle
{1 \over \langle u|G^{(+)}_v(\omega, \varepsilon)|u \rangle}
\nonumber \\
& & \times \langle u| \left[\,1+G^{(+)}_0(\omega)
T_v(\omega, \varepsilon)\,\right] \ ,
\label{form10}
\end{eqnarray}
where $T_v(\omega, \varepsilon)$ is defined by
\begin{eqnarray}
T_v(\omega, \varepsilon)
=v(\varepsilon)+v(\varepsilon)\,G^{(+)}_0(\omega)
\,T_v(\omega, \varepsilon) \ ,
\label{form11}
\end{eqnarray}
and $G^{(+)}_v(\omega, \varepsilon)$ is the corresponding full
Green function.

The introduction of the parameter $\varepsilon$ in \eq{form9} is not 
strange if we consider the practical method to solve equations
like Eqs.\,(\ref{form1}) and (\ref{form6}) with a redundant solution
$|u\rangle$.
In these equations we are actually solving $\Lambda \chi$ or
$\Lambda \psi$. The original Saito's suggestion \cite{SA68,SA77} for
solving the OCM equation \eq{form6} is to assume $\Lambda \psi=\psi$
or $\langle u|\psi \rangle=0$ and solve
\begin{equation}
\omega \psi=\Lambda \left( H_0+V_{\rm D}\right) \Lambda \psi \ \ .
\label{form12}
\end{equation}
In fact, the solution of \eq{form12} is the redundancy-free solution
of \eq{form6} when $\omega \neq 0$.
Generally speaking, the trivial solution $\vert u \rangle$ needs not
be a zero-energy solution. One can also move this exceptional
energy to an arbitrary (usually positive) value $\varepsilon$.
This can be achieved by simply adding 
$\varepsilon |u\rangle \langle u|\psi \rangle$ term in the right-hand
side of \eq{form12}:
\begin{equation}
\omega \psi=\Lambda (H_0+V_{\rm D})\Lambda \psi
+\varepsilon |u\rangle \langle u|\psi \rangle \ \ .
\label{form13}
\end{equation}
One can take the same process in the RGM equation \eq{form1}.
We start from \eq{form4} and change $\omega$ in $v(\omega)$
to $\varepsilon$. This is permissible since the energy dependence
of the $v(\varepsilon)$ term is usually very weak due to
the structure $\Lambda K \Lambda$. For example, this term
vanishes completely for simple systems like the two di-neutron
system. \cite{TRGM} The energy dependence of the RGM interaction
in the allowed space is later taken into account
by a self-consistency condition.
Similarly to \eq{form13}, we set up with the equation
\begin{equation}
\omega \chi=\Lambda \left[\,H_0+v(\varepsilon)\,\right] \Lambda \chi
+\varepsilon |u\rangle \langle u|\chi \rangle \ \ .
\label{form14}
\end{equation}
If we use \eq{form7}, we can easily prove that this equation is
nothing but
\begin{eqnarray}
\left[\,\omega -H_0-\VRGM\,\right] \chi=0\ ,
\label{form15}
\end{eqnarray}
which no longer has the trivial solution $|u\rangle$ except
for $\omega = \varepsilon$.

A motivation to use $\widetilde{T}(\omega, \varepsilon)$ in
\eq{form10} for
the Faddeev equation comes from the complete equivalence
between the Faddeev equation and the 3-body equation interacting
via $v(\varepsilon)$ in the allowed model space. \cite{TRGM}
Namely, for a system composed of three identical spinless particles,
we can prove the equivalence between
\begin{eqnarray}
P \left[\,E-H_0-\VRGMA-\VRGMB-\VRGMC\,\right] P \Psi=0
\label{form16}
\end{eqnarray}
and
\begin{eqnarray}
\psi_\alpha=G^{(+)}_0(E)\,\TTE \,(\psi_\beta+\psi_\gamma)\ ,
\label{form17}
\end{eqnarray}
where a common self-consistency condition
\begin{eqnarray}
\varepsilon_\alpha=\langle P \Psi|
\,h_\alpha+\VRGMA\,|P \Psi \rangle
/\langle P \Psi| P \Psi \rangle ,
\label{form18}
\end{eqnarray}
is imposed.
In \eq{form16}, $H_0$ is the three-body kinetic energy operator
in the c.m. system, $\VRGMA$ represents
the RGM kernel \eq{form2} for
the $\alpha$ pair, and  $P$ is the projection operator
onto the [3]-symmetric Pauli-allowed space,
as defined in Refs.\,\cite{TRGM} and \cite{HO74}.
In the Faddeev equation \eq{form17},
$\TTE$ is essentially the non-singular
RGM $T$-matrix $\widetilde{T}(\omega, \varepsilon)$ defined
through \eq{form10}:
\begin{equation}
\TTE=\widetilde{T}_\alpha(E-h_{\bar{\alpha}},
\varepsilon_\alpha)\ ,
\label{form19}
\end{equation}
where $h_{\bar{\alpha}}$ is the relative kinetic-energy operator
between the $\alpha$-pair and the third particle.
A nice point of the Faddeev equation \eq{form17} is that the total
wave function $\Psi$, constructed from the three Faddeev components,
$\psi_\alpha$, $\psi_\beta$ and $\psi_\gamma$ is automatically
orthogonal to the Pauli-forbidden state in each pair:
\begin{equation}
\langle u_\alpha |\Psi \rangle=\langle u_\alpha | P \Psi \rangle
=\langle u_\alpha |\psi_\alpha+\psi_\beta+\psi_\gamma \rangle=0 \ .
\label{form20}
\end{equation}
This is because of the orthogonality property
\begin{eqnarray}
\langle u| \left[\,1+G^{(+)}_0(\omega)
\widetilde{T}(\omega, \varepsilon)
\,\right]=\left[\, 1+\widetilde{T}(\omega, \varepsilon)
G^{(+)}_0(\omega)\,\right] |u \rangle=0 \ ,
\label{form21}
\end{eqnarray}
which is derived from the formal solution in \eq{form10}.

\subsection{Application to the 3-cluster OCM}

The above discussion on the 3-cluster systems interacting
via pairwise 2-cluster RGM kernels can be straightforwardly extended
to the ordinary 3-cluster OCM interacting
via simple energy-independent local potentials.
One only needs to modify
\begin{eqnarray}
v(\varepsilon) \longrightarrow v = \Lambda V_{\rm D} \Lambda \ .
\label{form22}
\end{eqnarray}
From Eqs.\,(\ref{form7}) and (\ref{form8}), the interaction term
for the 2-cluster OCM equation, $\VOCM$, turns out to be
\begin{eqnarray}
\VRGM \longrightarrow \VOCM & = & V(\varepsilon)+v
=(\varepsilon-H_0)-\Lambda (\varepsilon-H_0-V_{\rm D}) \Lambda
\nonumber \\
& = & \varepsilon |u\rangle \langle u|
+\Lambda \left( H_0+V_{\rm D} \right)
\Lambda-H_0 \ .
\label{form23}
\end{eqnarray}
The full $T$-matrix of $\VOCM$ is defined through
\begin{equation}
T(\omega,\varepsilon)=\VOCM+\VOCM G^{(+)}_0(\omega)
T(\omega,\varepsilon) \ .
\label{form24}
\end{equation}
The formal expression of $T(\omega,\varepsilon)$ is very similar to
\eq{form10}, but this time the ``OCM $T$-matrix'',
$\widetilde{T}(\omega)$, does not involve
the $\varepsilon$-dependence.
Namely, the $\varepsilon$-dependence appears only through the
last term in $T(\omega,\varepsilon)$.
Now we can write down two equivalent equations corresponding to
Eqs.\,(\ref{form16}) and (\ref{form17}):
\begin{eqnarray}
P \left[\,E-H_0-V^{\rm D}_\alpha-V^{\rm D}_\beta-V^{\rm D}_\gamma
\,\right] P \Psi=0
\label{form25}
\end{eqnarray}
and
\begin{eqnarray}
\psi_\alpha=G^{(+)}_0(E)\,\widetilde{T}^{(3)}_\alpha(E)
\,(\psi_\beta+\psi_\gamma)\ .
\label{form26}
\end{eqnarray}
This time, we do not need the self-consistency condition \eq{form18},
and \eq{form19} becomes
\begin{equation}
\widetilde{T}^{(3)}_\alpha (E)=\widetilde{T}_\alpha
(E-h_{\bar{\alpha}}) \ .
\label{form27}
\end{equation}
Since the orthogonality condition in \eq{form21} is still valid
for $\widetilde{T}(\omega)$, we can prove that the solution
of \eq{form26} satisfies the orthogonality
of the total wave function, \eq{form20}.

If the Pauli-forbidden state $\vert u \rangle$ is
a real bound state of $V_{\rm D}$,
our expression for $\VOCM$ in \eq{form23} is
further simplified. We assume that $|u\rangle=|u_B \rangle$ is
the bound-state wave function, satisfying
\begin{equation}
\left( \varepsilon_B -H_0-V_{\rm D} \right) \vert u_B \rangle =0
 \ \ ,
\label{form28}
\end{equation}
with $\varepsilon_B$ ($< 0$) being the bound-state energy.
Here, $H_0$ is the 2-cluster kinetic-energy operator,
and only one bound state is assumed to exist.
Then, one can easily show that $\VOCM$ is reduced to
\begin{equation}
\VOCM=V_{\rm D}+(\varepsilon - \varepsilon_B)
\vert u_B \rangle \langle u_B \vert \ \ ,
\label{form29}
\end{equation}
which is nothing but the Kukulin's pseudo-potential
\begin{equation}
V^{\rm ps}=V_{\rm D}+\lambda \vert u_B \rangle \langle u_B \vert
\label{form30}
\end{equation}
with $\lambda=\varepsilon - \varepsilon_B$.
By using the general formula given in Appendix, we can find that
the $T$-matrix defined through \eq{form24} is given by 
\begin{eqnarray}
& & T(\omega, \varepsilon)=T_{\rm D}(\omega)
+\left[\,1+T_{\rm D}(\omega) G_0(\omega)\,\right] \vert u_B \rangle
{1 \over {1 \over \varepsilon - \varepsilon_B}-\langle u_B \vert
G^{(+)}_D(\omega) \vert u_B \rangle} \nonumber \\
& & \times \langle u_B \vert \left[\,1+G_0(\omega)
T_{\rm D}(\omega)\,\right]
\ ,
\label{form31}
\end{eqnarray}
where
\begin{equation}
T_{\rm D}(\omega)=V_{\rm D}+V_{\rm D}\,G^{(+)}_0(\omega)
\,T_{\rm D}(\omega) \ \ ,
\label{form32}
\end{equation}
and
\begin{equation}
G^{(+)}_{\rm D}(\omega)=\left[\,\omega-H_0-V_{\rm D}
+i 0\,\right]^{-1}
=G_0(\omega)+G_0(\omega)\,T_{\rm D}(\omega)\,G_0(\omega) \ .
\label{form33}
\end{equation}
On the other hand, $(\omega-H_0-V_{\rm D})\vert u_B \rangle
=(\omega-\varepsilon_B)\vert u_B \rangle$ yields
\begin{equation}
G^{(+)}_{\rm D}(\omega) \vert u_B \rangle
= {1 \over \omega-\varepsilon_B}
\,\vert u_B \rangle \ ,
\label{form34}
\end{equation}
if $\omega \neq \varepsilon_B$.
By using \eq{form34}, the second term of \eq{form31} in the
right-hand side is greatly simplified through the relationship like
\begin{eqnarray}
\left[\,1+T_{\rm D}(\omega) G_0(\omega)\,\right] \vert u_B \rangle
=(\omega-H_0) G^{(+)}_{\rm D}(\omega) \vert u_B \rangle
=(\omega-H_0) \vert u_B \rangle {1 \over \omega-\varepsilon_B}.
\label{form35}
\end{eqnarray}
We find
\begin{eqnarray}
& & T(\omega, \varepsilon)=T_{\rm D}(\omega)
-(\omega-H_0) \vert u_B \rangle {1 \over \omega-\varepsilon_B}
\langle u_B \vert (\omega-H_0) \nonumber \\
& & +(\omega-H_0) \vert u_B \rangle {1 \over \omega-\varepsilon}
\langle u_B \vert (\omega-H_0) \ .
\label{form36}
\end{eqnarray}
If we define $\widetilde{T}(\omega)$ through
\begin{eqnarray}
T(\omega, \varepsilon)=\widetilde{T}(\omega)
+(\omega-H_0) \vert u_B \rangle {1 \over \omega-\varepsilon}
\langle u_B \vert (\omega-H_0) \ ,
\label{form37}
\end{eqnarray}
we obtain
\begin{eqnarray}
\widetilde{T}(\omega)=T_{\rm D}(\omega)
-(\omega-H_0) \vert u_B \rangle {1 \over \omega-\varepsilon_B}
\langle u_B \vert (\omega-H_0) \ ,
\label{form38}
\end{eqnarray}
or
\begin{eqnarray}
\widetilde{T}(\omega)=\lim_{\varepsilon \rightarrow \infty}
T(\omega, \varepsilon) \ . 
\label{form39}
\end{eqnarray}

Adding the separable term to $V_{\rm D}$ in \eq{form29} removes
the bound-state pole of the $T$-matrix $T_{\rm D}(\omega)$ and
moves it to the positive energy $\varepsilon$ in \eq{form37}.
In $\widetilde{T}(\omega)$ in \eq{form38}, this
positive energy pole is even removed. In order to see this, we
use the spectral decomposition of $G^{(+)}_{\rm D}(\omega)$:
\begin{eqnarray}
& & G^{(+)}_{\rm D}(\omega)=\left[\,\omega-H_0
-V_{\rm D}+i 0\,\right]^{-1}
\nonumber \\
& & =\vert u_B \rangle {1 \over \omega-\varepsilon_B+i0}
\langle u_B \vert
+ \int d \bk \vert \phi^{(+)}_{\bk} \rangle
{1 \over \omega-{\hbar^2 \over 2\mu}\bk^2+i0}
\langle \phi^{(+)}_{\bk} \vert \ .
\label{form40}
\end{eqnarray}
Using this in $T_{\rm D}(\omega)=V_{\rm D}+V_{\rm D}
\,G^{(+)}_{\rm D}(\omega)\,V_{\rm D}$ and in \eq{form38}, we find
\begin{eqnarray}
\widetilde{T}(\omega) & = & \Lambda V_{\rm D} \Lambda
-\vert u_B \rangle \langle u_B \vert (\omega - H_0)
\vert u_B \rangle \langle u_B \vert \nonumber \\
& & +\int d \bk V_{\rm D} \vert \phi^{(+)}_{\bk} \rangle
{1 \over \omega-{\hbar^2 \over 2\mu}\bk^2+i0}
\langle \phi^{(+)}_{\bk} \vert V_{\rm D} \ ,
\label{form41}
\end{eqnarray}
which has no singularities if $\omega < 0$.

Let us consider a three-body system composed of three identical
spinless particles with the mass $M$.
One of such examples is the $3\alpha$ system interacting via
the BFW potential, discussed in the next section.
The pairwise interaction is assumed to be
the Kukulin's pseudo-potential $V^{\rm ps}$ with
$\lambda=\varepsilon - \varepsilon_B$ in \eq{form30}.
The three-body Schr{\"o}dinger equation reads
\begin{equation}
\left( E-H_0-V^{\rm ps}_\alpha-V^{\rm ps}_\beta
-V^{\rm ps}_\gamma \right)
\Psi=0 \ \ .
\label{form42}
\end{equation}
This equation is transformed to the Faddeev equation by the
standard procedure:
\begin{equation}
\psi_\alpha=G_0\,T_\alpha
\,\left( \psi_\beta+\psi_\gamma \right) \ \ ,
\label{form43}
\end{equation}
where $G_0=G^{(+)}_0(E)$ with $E<0$, $T_\alpha
=T_\alpha (E-h_{\bar \alpha})$,
and $\psi_\alpha$ etc. are the three Faddeev components
yielding the total wave function $\Psi=\psi_\alpha+\psi_\beta
+\psi_\gamma$.
Using \eq{form37}, we find
\begin{eqnarray}
\psi_\alpha & = & G_0 \left( \widetilde{T}_\alpha
+(E-H_0)\,\vert u^B_\alpha \rangle {1 \over E-{3 \over 4}
{\hbar^2 \over M}{\bq_\alpha}^2-\varepsilon}
\langle u^B_\alpha \vert (E-H_0) \right)
\left( \psi_\beta+\psi_\gamma \right) \nonumber \\
& = & G_0 \widetilde{T}_\alpha \left( \psi_\beta+\psi_\gamma \right)
+\vert u^B_\alpha \rangle {1 \over E-{3 \over 4}{\hbar^2 \over M}
{\bq_\alpha}^2-\varepsilon} \langle u^B_\alpha \vert E-H_0 \vert 
\psi_\beta+\psi_\gamma \rangle \ .
\label{form44}
\end{eqnarray}
Here, $\bq$ is the momentum Jacobi coordinate between a pair and
the third particle. 
If we multiply \eq{form44} by $\langle u^B_\alpha \vert$ from
the left-hand side, we obtain
\begin{eqnarray}
\langle u^B_\alpha \vert \psi_\alpha \rangle
= -\langle u^B_\alpha \vert \psi_\beta+\psi_\gamma \rangle
+{1 \over E-{3 \over 4}{\hbar^2 \over M}
{\bq_\alpha}^2-\varepsilon} \langle u^B_\alpha \vert E-H_0 \vert 
\psi_\beta+\psi_\gamma \rangle \ ,\qquad
\label{form45}
\end{eqnarray}
or
\begin{equation}
\langle u^B_\alpha \vert \Psi \rangle
={1 \over E-{3 \over 4}{\hbar^2 \over M}
{\bq_\alpha}^2-\varepsilon} \langle u^B_\alpha \vert E-H_0 \vert 
\psi_\beta+\psi_\gamma \rangle \ .
\label{form46}
\end{equation}
If we take the limit $\lambda \rightarrow \infty$ or 
$\varepsilon \rightarrow \infty$, we obtain
\begin{equation}
\psi_\alpha=G_0 \widetilde{T}_\alpha
\left( \psi_\beta+\psi_\gamma \right) \ ,
\label{form47}
\end{equation}
and
\begin{equation}
\langle u^B_\alpha \vert \Psi \rangle=0 \ .
\label{form48}
\end{equation}
Since $\widetilde{T}$ does not depend on $\varepsilon$ or $\lambda$,
we can achieve the solution of \eq{form42} with $\lambda
\rightarrow \infty$,
by solving \eq{form47} without any limiting procedure.
On the other hand, we have proven that \eq{form47} is
equivalent to the Horiuchi's OCM \eq{form25}, using $V_{\rm D}$.
We can thus prove the equivalence between \eq{form42} with
$\lambda \rightarrow \infty$ and \eq{form25};
namely, the equivalence between the Kukulin's OCM and
the Horiuchi's OCM,
when the Pauli-forbidden state $\vert u \rangle$ is
the exact eigen-state for the local potential $V_{\rm D}$. 
In fact, this equivalence is also valid even
when the $\vert u \rangle$ is
not the eigen-state of $V_{\rm D}$
but the h.o. Pauli-forbidden state of the normalization
kernel $K$, which is proved in the next subsection.
 
For solving the $3\alpha$ Faddeev equation, it is important to note
that there exist some trivial solutions related to the orthogonality
condition \eq{form21} for the $\widetilde{T}$ $T$-matrix.
As is discussed in Ref.\,\cite{RED} in detail,
the eigen-value solutions of the rearrangement matrix $S$ among
the three different types of the Jacobi coordinates
with the eigen-value $\tau=-1$ are the redundant solutions
of the Faddeev equation Eqs.\,(\ref{form17}) or (\ref{form26}).
For the h.o. Pauli-forbidden state $|u\rangle$,
this eigen-value equation reads 
\begin{equation}
\langle u|S|u f^\tau \rangle=\tau |f^\tau \rangle
\qquad \hbox{with} \quad \tau=-1 \ .
\label{red3}
\end{equation}
We find that the Faddeev component
\begin{equation}
\psi^\tau_0=G_0 |u f^\tau \rangle \qquad \hbox{with} \quad \tau=-1
\label{red4}
\end{equation}
is a trivial solution with the total wave
function $\Psi^\tau_0=(1+S)\psi^\tau_0=0$ for $\tau=-1$.
The non-zero $\psi^\tau_0$ is
the [21]-symmetric function with respect to the permutation of
the $3\alpha$ particles. We have two such trivial solutions for
the $3\alpha$ system with the total angular momentum $L=0$.
For this reason, the Faddeev
equation (\ref{form17}) or (\ref{form26}) should
be modified to\footnote{The inverse of the matrix elements
in the last terms of \protect\eq{red5} and \protect\eq{red10}
should be understood as the matrix inverse.}
\begin{eqnarray}
\lambda \psi=\left[ G_0 \widetilde{T} S
- \sum_{\tau=-1} G_0|u f^\tau\rangle
{1 \over \langle u f^\tau|G_0|u f^\tau\rangle}
\langle u f^\tau|\,\right]\psi\ ,
\label{red5}
\end{eqnarray}
in order to find a unique solution with $\lambda=1$.
The solution of \eq{red5} with $\lambda=1$ automatically satisfies
\begin{eqnarray}
\psi=G_0 \widetilde{T} S \psi\ ,\quad
\langle u |(1+S)|\psi\rangle=0\ ,\quad
\langle u f^\tau|\psi \rangle=0 \quad \hbox{for} \quad \tau=-1\ .
\label{red6}
\end{eqnarray}

When the bound-state solution $|u_B\rangle$ is used for $|u\rangle$,
these $\tau=-1$ eigen-values of \eq{red3} are no longer
exactly $\tau=-1$, but move to the $\tau > -1$ values.
The [3]-symmetric basis states constructed from 
\begin{equation}
\Psi_0={1 \over \sqrt{3(1+\tau)}}(1+S)|u_B f^\tau \rangle
\qquad \hbox{with} \quad \tau \sim -1
\label{red7}
\end{equation}
involve some of the important shell-model like components
such as $|[3](04)\rangle$ etc.\,\cite{FU03}
If these configurations are excluded,
one cannot describe a compact shell-model like structure
of $\hbox{}^{12}\hbox{C}$.
The original $3\alpha$ OCM equation should, therefore,
be formulated by using 
\begin{equation}
\widetilde{P}=|\Psi_0 \rangle \langle \Psi_0|+P \ \ ,
\label{red8}
\end{equation}
instead of $P$ in \eq{form25}.
Since $|\Psi_0 \rangle$ involves a small admixture of the
redundant components as
\begin{equation}
\langle u_B|\Psi_0\rangle=\sqrt{{1+\tau \over 3}}~|f^\tau \rangle
\qquad \hbox{with} \quad \tau \sim -1\ \ ,
\label{red9}
\end{equation}
the ground-state solution of the $3\alpha$ system with the
dominant $|\Psi_0 \rangle$ components naturally involves
a small admixture of the redundant components.
We can also formulate an equivalent Faddeev equation to this
modified $3\alpha$ OCM equation with $\widetilde{P}$,
which has a slightly different form from \eq{red5}:
\begin{eqnarray}
\lambda \psi=\left[ G_0 \widetilde{T} S
+\sum_{\tau \sim -1} |u f^\tau\rangle
{1 \over \langle u f^\tau|E-H_0-V_{\rm D}|u f^\tau\rangle}
\langle u f^\tau|(E-H_0)S\,\right]\psi\ .
\label{red10}
\end{eqnarray}
The derivation of this equation and a detailed discussion
of the almost redundant components of the Faddeev equation will be
given elsewhere.\,\cite{FU03}

\subsection{Equivalence between pairwise orthogonality conditions
and the method of orthogonalizing pseudo-potentials
in the 3-cluster systems}

In this subsection, we will prove the equivalence between
the Horiuchi's OCM and the Kukulin's OCM, even when the
Pauli-forbidden state $\vert u \rangle$ is not the eigen-state
of the pairwise potential $V_{\rm D}$,
but the original h.o. Pauli-forbidden state of $K$.
The essential point is that the OCM $T$-matrix
$\widetilde{T}(\omega)$
defined through $T(\omega,\varepsilon)$ in \eq{form24} is
nothing but the $\widetilde{T}$-matrix generated from
the pseudo-potential $V^{\rm ps}$ in \eq{form30}
with $|u_B\rangle \rightarrow |u\rangle$ and $\lambda
\rightarrow \infty$.

Let us assume that $|u\rangle$ is {\em not} the eigen-state
of $V_{\rm D}$, but the h.o. eigen-state of $K$ with the
eigen-value $\gamma=1$.
The $T$-matrix generated from $V^{\rm ps}$
with $\lambda \rightarrow \infty$ is, from \eq{form31},
\begin{eqnarray}
\widetilde{T}(\omega)
=T_{\rm D}(\omega)-(\omega -H_0) G_{\rm D}(\omega)
\vert u \rangle {1 \over \langle u \vert G_D(\omega) \vert u \rangle}
\langle u \vert G_D(\omega) (\omega -H_0) \ , 
\label{form50}
\end{eqnarray}
where $T_{\rm D}(\omega)$ and $G_{\rm D}(\omega)$ are
the $T$-matrix and the full Green function of $V_{\rm D}$,
respectively.
On the other hand, we separate $V^{\rm OCM}(\varepsilon)$
in \eq{form23} as
\begin{eqnarray}
V^{\rm OCM}(\varepsilon)
=(\varepsilon-H_0)-\Lambda (\varepsilon-H_0-V_{\rm D})\Lambda
=V_1+\lambda_1 |u\rangle \langle u|
\label{form51}
\end{eqnarray}
with
\begin{eqnarray}
& & V_1=V_{\rm D}-(H_0+V_{\rm D}) \vert u \rangle \langle u \vert
-\vert u \rangle \langle u \vert (H_0+V_{\rm D}) \ \ , \nonumber \\ 
& & \lambda_1=\varepsilon+\langle u \vert H_0+V_{\rm D}
\vert u \rangle
\ \ .
\label{form52}
\end{eqnarray}
Then, the general formula in Appendix gives the $T$-matrix as
\begin{eqnarray}
& & T(\omega, \varepsilon)=t_1+(1+t_1 G_0)\vert u\rangle
{1 \over {\lambda_1}^{-1}-\langle u \vert G_0+G_0 t_1 G_0
\vert u \rangle}
\langle u \vert (1+G_0 t_1) \nonumber \\
& & = t_1+(\omega -H_0) G_1 \vert u \rangle
{1 \over {\lambda_1}^{-1}-\langle u \vert G_1 \vert u \rangle}
\langle u \vert G_1 (\omega -H_0) \ ,
\label{form53}
\end{eqnarray}
where $t_1$ and $G_1$ are the $T$-matrix and the full Green function
of $V_1$.
Here we take $\varepsilon \rightarrow \infty$
($\lambda_1 \rightarrow \infty$) and obtain
\begin{eqnarray}
\widetilde{T}(\omega)=
\lim_{\varepsilon \rightarrow \infty} T(\omega, \varepsilon)
=t_1-(\omega -H_0) G_1 \vert u \rangle
{1 \over \langle u \vert G_1 \vert u \rangle}
\langle u \vert G_1 (\omega -H_0) \ .
\label{form54}
\end{eqnarray}
We note that $\omega-H_0-V_1$ is expressed as
\begin{eqnarray}
& & \omega-H_0-V_1=\omega-H_0-V_{\rm D}
+(H_0+V_{\rm D}) \vert u \rangle \langle u \vert
+\vert u \rangle \langle u \vert (H_0+V_{\rm D}) \nonumber \\
& & =\omega-\Lambda (H_0+V_{\rm D}) \Lambda
+\vert u \rangle \langle u \vert (H_0+V_{\rm D})
\vert u \rangle \langle u \vert \nonumber \\
& & =\Lambda (\omega -H_0-V_{\rm D}) \Lambda
+\vert u \rangle \left[\,\omega +\langle u \vert (H_0+V_{\rm D})
\vert u \rangle\,\right] \langle u \vert \ .
\label{form55}
\end{eqnarray}
Then, we find that $G_1=(\omega-H_0-V_1)^{-1}$ is expressed
as
\begin{eqnarray}
G_1=G_\Lambda+\vert u \rangle {1 \over \omega
+\langle u \vert H_0+V_{\rm D} \vert u \rangle}
\langle u \vert \ ,
\label{form56}
\end{eqnarray}
where $G_\Lambda$ is defined by
\begin{eqnarray}
G_\Lambda=G_{\rm D}-G_{\rm D} \vert u \rangle
{1 \over \langle u \vert G_{\rm D} \vert u \rangle}
\langle u \vert G_{\rm D} \ ,
\label{form57}
\end{eqnarray}
with $G_{\rm D}=(\omega-H_0-V_{\rm D})^{-1}$, and
satisfies
\begin{eqnarray}
\Lambda (\omega-H_0-V_{\rm D}) \Lambda\,G_\Lambda
=G_\Lambda\,\Lambda (\omega-H_0-V_{\rm D}) \Lambda = \Lambda \ .
\label{form58}
\end{eqnarray}
Since we have
\begin{eqnarray}
G_1 \vert u \rangle
=\vert u \rangle {1 \over \omega
+\langle u \vert H_0+V_{\rm D} \vert u \rangle}
\qquad \hbox{etc.} \ ,
\label{form59}
\end{eqnarray}
\eq{form54} becomes
\begin{eqnarray}
\widetilde{T}(\omega)=t_1-(\omega -H_0) \vert u \rangle
{1 \over  \omega+\langle u \vert H_0+V_{\rm D} \vert u \rangle}
\langle u \vert (\omega -H_0) \ .
\label{form60}
\end{eqnarray}
On the other hand, $G_1=G_0+G_0 t_1 G_0$ yields
\begin{eqnarray}
& & t_1=(\omega -H_0) G_1 (\omega -H_0)-(\omega -H_0)
= (\omega -H_0) G_\Lambda (\omega -H_0)-(\omega -H_0) \nonumber \\
& & +(\omega -H_0) \vert u \rangle
{1 \over  \omega+\langle u \vert H_0+V_{\rm D} \vert u \rangle}
\langle u \vert (\omega -H_0) \ ,
\label{form61}
\end{eqnarray}
where the last term in the right-hand side cancels with the
last term of \eq{form60}. Thus, we find
\begin{eqnarray}
& & \widetilde{T}(\omega)
=(\omega -H_0) G_\Lambda (\omega -H_0)-(\omega -H_0) \nonumber \\
& & =(\omega -H_0) G_{\rm D} (\omega -H_0)-(\omega -H_0)
-(\omega-H_0) G_{\rm D} \vert u \rangle
{1 \over \langle u \vert G_{\rm D} \vert u \rangle}
\langle u \vert G_{\rm D}(\omega-H_0) \nonumber \\
& & = T_{\rm D}-(\omega-H_0) G_{\rm D} \vert u \rangle
{1 \over \langle u \vert G_{\rm D} \vert u \rangle}
\langle u \vert G_{\rm D}(\omega-H_0) \ ,
\label{form62}
\end{eqnarray}
which is nothing but \eq{form50}.
We can also prove that the bound state pole of $V_{\rm D}$ is
eliminated from \eq{form62}, by using the spectral decomposition
of $G_{\rm D}$ in \eq{form40}.
This proves the equivalence between the Kukulin's OCM and
our Faddeev OCM, hence the equivalence between the Kukulin's OCM
and the Horiuchi's OCM for the h.o. Pauli-forbidden
state $|u\rangle$.

\section{\mib{3\alpha} OCM for the BFW \mib{\alpha \alpha} potential}

\begin{table}[b]
\caption{Two-$\alpha$ ($2\alpha$) bound-state energies
of the BFW potential with and without the Coulomb force.
The heading ``momen. rep.'' stands for
the $2\alpha$ bound-state energies calculated
in the $3\alpha$ Faddeev code,
by using the Born kernel including the cut-off Coulomb force
with $R_C=10$ fm. See the text for the momentum discretization
points. RKG is the results by the Runge-Kutta-Gill method
in the coordinate representation (full Coulomb).
The Coulomb contribution is also shown in the rows ``Coulomb cont.''
}
\label{table1}
\begin{center}
\renewcommand{\arraystretch}{1.4}
\setlength{\tabcolsep}{4mm}
\begin{tabular}{cccc}
\hline
momen. rep.     & $0s$ & $1s$ & $0d$ \\
\hline
with Coulomb    & $-72.6257$ & $-25.6186$ & $-22.0005$ \\
without Coulomb & $-76.7051$ & $-28.7325$ & $-25.0908$ \\
Coulomb cont.   &  4.0794    &   3.1139   &   3.0903   \\
\hline
RKG             & $0s$ & $1s$ & $0d$ \\
\hline
with Coulomb    & $-72.6255$ & $-25.6186$ & $-22.0005$ \\
without Coulomb & $-76.7050$ & $-28.7325$ & $-25.0908$ \\
Coulomb cont.   &  4.0795    &   3.1139   &   3.0903   \\
\hline
\end{tabular}
\end{center}
\end{table}

As an example, we consider the $3\alpha$ system interacting via the
deep $\alpha \alpha$ potential
\begin{equation}
V^{\rm BFW}(r)=-122.6225~e^{-0.22~r^2}
+4e^2\,\erf(0.75~r)/r  \qquad (\hbox{MeV})
\label{res1}
\end{equation}
with $r$, being the relative coordinate between $2\alpha$'s in fm.
In \eq{res1}, $\erf(x)=(2/\sqrt{\pi}) \int^x_0 e^{-t^2} dt$ is
the error function.
The Pauli principle between $2\alpha$ particles
are taken into account
in terms of the bound states of $V^{\rm BFW}(r)$.
Since the Coulomb force is not exactly treated
in the Faddeev formalism,
we replace the folded Coulomb term of \eq{res1} with
the cut-off Coulomb force $v^{\rm C}(r)
=(4e^2/r)\,\theta(R_C-r)$ with $R_C=10$ fm, introduced 
at the nucleon level. Here $\theta(x)$ is the Heaviside step
function. For this type of the Coulomb force between protons,
the $\alpha \alpha$ Coulomb potential in \eq{res1} is replaced with
\begin{eqnarray}
V^{\rm C}_{\alpha \alpha}(r) & = & 4e^2\,\left\{\,\erf(0.75~r)
\right.
\nonumber \\
& & \left. -(1/2)\left[\,\erf(0.75~(r+R_C))
+\erf(0.75~(r-R_C))\,\right]
\,\right\}/r \ .
\label{res2}
\end{eqnarray}
In the following, we use $\hbar^2/M_\alpha=10.4465
~\hbox{MeV}\cdot \hbox{fm}^2$ and $e^2=1.44~\hbox{MeV}
\cdot \hbox{fm}$ for
the comparison with Ref.\,\cite{TB03}, unless otherwise specified. 
Table \ref{table1} shows the bound-state energies obtained by
diagonalizing the negative-energy $T$-matrix
in the momentum representation. We have two bound states,
$(0s)$ and $(1s)$,
in the relative $S$-state and one bound state, $(0d)$,
in the $D$-state, when the Coulomb force is included.
The relative momentum $p$ is descritized
using the Gauss-Legendre $n_1$-point quadrature formula
for each of the four intervals of 0 - 1 $\hbox{fm}^{-1}$,
1 - 3 $\hbox{fm}^{-1}$, 3 - 6 $\hbox{fm}^{-1}$,
and 6 - 12 $\hbox{fm}^{-1}$.
The large value of $p$ beyond $p_0=12~\hbox{fm}^{-1}$ is also taken
into account by using the Gauss-Legendre $n_3$-point quadrature
formula through the mapping $p=p_0+{\rm tan}\left\{\pi
(1+x)/4\right\}$.\footnote{These $n_3$ points
for $p$ are not included for solving the Faddeev
equation (\protect\ref{red5}) or (\protect\ref{red10}),
since these cause a numerical inaccuracy for the interpolation.}
We choose $n_1=15$ and $n_3=5$, so that 65 points are
used for solving the bound-state wave functions
and the necessary $T$-matrix for solving the Faddeev equation.
The second Jacobi coordinate $q$
is also descritized with the Gauss-Legendre $n_2$-point
quadrature formula with $n_2=10$ in the similar way to $p$,
but in this case choosing only three major dividing points
with $q=1~\hbox{fm}^{-1},
3~\hbox{fm}^{-1}$ and 6 $\hbox{fm}^{-1}$ is good enough.
The Gauss-Legendre $n_3$-point quadrature formula
with $q=q_0+{\rm tan}\left\{\pi(1+x)/4\right\}$ is also applied
to $q \geq q_0=6~\hbox{fm}^{-1}$.
In Table \ref{table1}, the result of the coordinate-space calculation
using the Runge-Kutta-Gill method is also shown for comparison.
In this case, we use the full Coulomb force in \eq{res1}.
We find that the choice $R_C=10$ fm
in the momentum-representation
calculation is accurate enough with the error less than 1 keV.
This is also true even in the $3\alpha$ Faddeev calculations.
The $\alpha \alpha$ phase shifts by the BFW potential in \eq{res1}
is very well reproduced,
as is shown in the original paper \cite{BF77}. 

\begin{table}[t]
\caption{Results of the $3\alpha$ Faddeev calculations
for the ground (${0_1}^+$) and excited (${0_2}^+$) $0^+$ states
of $\hbox{}^{12}\hbox{C}$, by using the BFW potential
and the cut-off Coulomb force with $R_C=10$ fm.
The bound-state (b.s.) wave functions of the BFW potential,
$|u_B\rangle$, are used for the Pauli-forbidden states $|u\rangle$.
Partial waves up to $\ell_{\rm max}$ are included in $2\alpha$ and
$(2\alpha)$-$\alpha$ relative motion.
The heading $n_1$-$n_2$-$n_3$ is the number of the momentum
discretization points (see the text for details);
``dim.'' stands for the full dimensionality
of the diagonalizing matrix for the Faddeev equation (i.e.,
$4n_1 (3n_2+n_3)(\ell_{\rm max}/2+1)$);
$\varepsilon_{2\alpha}$ is the expectation value
of the 2$\alpha$ Hamiltonian with respect
to the 3$\alpha$ bound-state solution;
$E_{3\alpha}$ is the 3$\alpha$ bound-state energy;
$c_{(04)}$ is the overlap amplitude
between the 3$\alpha$ bound-state
wave function and the $SU_3$ (04) configuration
with the h.o. width parameter $\nu=0.28125~\hbox{fm}^{-2}$;
and $\langle f|f \rangle$ in the last column is the squared norm
of the redundant components.
The results of the variational calculations,
using the translationally invariant h.o. basis
with the maximum h.o. quanta $N_{\rm max}=72$, is also shown
in the last rows for comparison.
}
\label{table2}
\begin{center}
\renewcommand{\arraystretch}{1.4}
\setlength{\tabcolsep}{2mm}
\begin{tabular}{cccccccc}
\hline
$|u\rangle$ & $\ell_{\rm max}$ & $n_1$-$n_2$-$n_3$ & dim.
& $\varepsilon_{2\alpha}$ (MeV) & $E_{3\alpha}$ (MeV) & $c_{(04)}$
& $\langle f|f \rangle$ \\
\hline
     &  4 & 15-10-5 &  6,300 & 14.610 & $-19.595$ & 0.9652 &
$2.7 \times 10^{-4}$ \\
b.s. &  6 & 15-10-5 &  8,400 & 14.485 & $-19.894$ & 0.9613 &
$2.7 \times 10^{-4}$ \\
${0_1}^+$ &  8 & 15-10-5 & 10,500 & 14.483 & $-19.897$ & 0.9612 &
$2.7 \times 10^{-4}$ \\ 
     & 10 & 15-10-5 & 12,600 & 14.483 & $-19.897$ & 0.9612 &
$2.7 \times 10^{-4}$ \\
\multicolumn{4}{c}{h.o.~variation} & 14.482 & $-19.897$ &  &
$2.7 \times 10^{-4}$ \\
\hline
     &  4 & 15-10-5 &  6,300 & 7.886 & $-0.370$ & 0.1302 &
$1.9 \times 10^{-6}$ \\
b.s. &  6 & 15-10-5 &  8,400 & 8.431 & $-0.485$ & 0.1410 &
$1.7 \times 10^{-6}$ \\
${0_2}^+$ &  8 & 15-10-5 & 10,500 & 8.500 & $-0.475$ & 0.1419 &
$1.7 \times 10^{-6}$ \\ 
     & 10 & 15-10-5 & 12,600 & 8.522 & $-0.471$ & 0.1422 &
$1.7 \times 10^{-6}$ \\
\multicolumn{4}{c}{h.o.~variation} & 9.950 & $-0.241$ &  &
$2.0 \times 10^{-6}$ \\
\hline
\end{tabular}
\end{center}
\end{table}
\begin{table}[hb]
\caption{The same as Table \protect\ref{table2}, but when
the Coulomb force is switched off. 
In the upper half, denoted by ``h.o.'', the results for the
h.o. Pauli-forbidden states $|u\rangle$ with the width
parameter $\nu=0.28125~\hbox{fm}^{-2}$ are shown.
}
\label{table3}
\begin{center}
\renewcommand{\arraystretch}{1.4}
\setlength{\tabcolsep}{2mm}
\begin{tabular}{cccccccc}
\hline
$|u\rangle$ & $\ell_{\rm max}$ & $n_1$-$n_2$-$n_3$ & dim.
& $\varepsilon_{2\alpha}$ (MeV) & $E_{3\alpha}$ (MeV) & $c_{(04)}$
& $\langle f|f \rangle$ \\
\hline
     &  4 & 15-10-5 &  6,300 & 12.405 & $-25.296$ & 0.9740 &
$1.3 \times 10^{-12}$ \\
h.o. &  6 & 15-10-5 &  8,400 & 12.262 & $-25.563$ & 0.9707 &
$1.3 \times 10^{-12}$ \\
${0_1}^+$ &  8 & 15-10-5 & 10,500 & 12.259 & $-25.565$ & 0.9706 &
$1.3 \times 10^{-12}$ \\
     & 10 & 15-10-5 & 12,600 & 12.259 & $-25.565$ & 0.9706 &
$1.3 \times 10^{-12}$ \\
\multicolumn{4}{c}{h.o.~variation} & 12.258 & $-25.565$ & 0.9706 & \\
\hline
     &  4 & 15-10-5 &  6,300 & 6.836 & $-6.181$ & 0.1343 &
$5.7 \times 10^{-12}$ \\
h.o. &  6 & 15-10-5 &  8,400 & 7.117 & $-6.404$ & 0.1423 &
$5.8 \times 10^{-12}$ \\
${0_2}^+$ &  8 & 15-10-5 & 10,500 & 7.125 & $-6.417$ & 0.1426 &
$5.8 \times 10^{-12}$ \\
      & 10 & 15-10-5 & 12,600 & 7.124 & $-6.418$ & 0.1427 &
$5.8 \times 10^{-12}$ \\
\multicolumn{4}{c}{h.o.~variation} & 7.165 & $-6.412$ & 0.1430 & \\
\hline
     &  4 & 15-10-5 &  6,300 & 12.722 & $-27.431$ & 0.9738 &
$2.6 \times 10^{-4}$ \\
b.s. &  6 & 15-10-5 &  8,400 & 12.582 & $-27.745$ & 0.9700 &
$2.6 \times 10^{-4}$ \\
${0_1}^+$ &  8 & 15-10-5 &  10,500 & 12.580 & $-27.748$ & 0.9700 &
$2.6 \times 10^{-4}$ \\
     & 10 & 15-10-5 & 12,600 & 12.580 & $-27.748$ & 0.9700 &
$2.6 \times 10^{-4}$ \\
\multicolumn{4}{c}{h.o.~variation} & 12.579 & $-27.748$ &    &
$2.6 \times 10^{-4}$ \\
\hline
     &  4 & 15-10-5 &  6,300 & 9.062 & $-5.731$ & 0.1121 &
$3.1 \times 10^{-6}$ \\
b.s. &  6 & 15-10-5 &  8,400 & 9.327 & $-6.060$ & 0.1213 &
$2.5 \times 10^{-6}$ \\
${0_2}^+$ &  8 & 15-10-5 & 10,500 & 9.329 & $-6.077$ & 0.1217 &
$2.5 \times 10^{-6}$ \\
     & 10 & 15-10-5 & 12,600 & 9.328 & $-6.078$ & 0.1217 &
$2.5 \times 10^{-6}$ \\
\multicolumn{4}{c}{h.o.~variation} & 9.363 & $-6.074$ &    &
$2.5 \times 10^{-6}$ \\
\hline
\end{tabular}
\end{center}
\end{table}

\begin{table}[ht]
\caption{Result of the 3$\alpha$ OCM with pairwise orthogonality
conditions (Horiuchi-type 3$\alpha$ OCM) for the $3\alpha$ ground state,
using the translationally invariant h.o. basis.
The BFW potential for $\alpha \alpha$ is used
with $\hbar^2/M_N=41.7860~\hbox{MeV}\cdot \hbox{fm}^2$
and $\nu=0.28125~\hbox{fm}^{-2}$. The Coulomb force is switched off.
The heading $N_{\rm max}$ is the maximum value of the
total h.o quanta included in the calculation;
$E_{2\alpha}$ is the lowest 2$\alpha$ diagonalization energy;
$\varepsilon_{2\alpha}$ is the expectation value
of the 2$\alpha$ Hamiltonian with respect
to the 3$\alpha$ ground-state solution;
$E_{3\alpha}$ is the 3$\alpha$ ground-state energy;
and $c_{(04)}$ is the overlap amplitude between
the 3$\alpha$ ground-state wave function
and the $SU_3$ (04) configuration.
The Faddeev result in Table \protect\ref{table3} with
the h.o. $|u\rangle$ is also shown in the bottom row for comparison.
}
\label{table4}
\begin{center}
\renewcommand{\arraystretch}{1.2}
\setlength{\tabcolsep}{4mm}
\begin{tabular}{ccccc}
\hline
$N_{\rm max}$ & $E_{2\alpha}$ (MeV) & $\varepsilon_{2\alpha}$ (MeV)
              & $E_{3\alpha}$ (MeV) & $c_{(04)}$ \\
\hline
4  &   9.81970  & $-$ & $-$ & $-$ \\
6  &   6.23729  & $-$ & $-$ & $-$ \\
8  &   2.34421  & 14.0067  & $-22.6177$ &    1   \\
10 &   0.85837  & 14.3470  & $-22.6745$ & 0.9993 \\
12 & $-0.11812$ & 12.9759  & $-25.2357$ & 0.9811 \\
14 & $-0.67019$ & 12.7748  & $-25.2768$ & 0.9797 \\
16 & $-1.03181$ & 12.4909  & $-25.5086$ & 0.9740 \\
18 & $-1.26800$ & 12.3918  & $-25.5244$ & 0.9729 \\
20 & $-1.43122$ & 12.3215  & $-25.5528$ & 0.9716 \\
30 & $-1.77280$ & 12.2608  & $-25.5649$ & 0.9707 \\
40 & $-1.86319$ & 12.2584  & $-25.5652$ & 0.9706 \\
50 & $-1.89424$ & 12.2583  & $-25.5652$ & 0.9706 \\
60 & $-1.90664$ & 12.2583  & $-25.5652$ & 0.9706 \\
\hline
Faddeev &       & 12.2594  & $-25.5653$ & 0.9706  \\
\hline
\end{tabular}
\end{center}
\end{table}

Table \ref{table2} shows the solutions of the $3\alpha$ Faddeev
equation (\ref{red10}) for the $L=0$ ground and excited states,
obtained by using the BFW potential and the cut-off Coulomb
force with $R_C=10$ fm. The demarcation ``b.s.'' implies
that the bound-state wave functions
of the BFW potential are used for the Pauli-forbidden states.
Partial waves up to $\ell_{\rm max}$ are included in $2\alpha$ and
$(2\alpha)$-$\alpha$ relative motion. The convergence of the ground 
state energy is very rapid
when we increase $\ell_{\rm max}$ from 4 to 10.
The inaccuracy in $\ell_{\rm max}=8$ is within 1 keV.
Table \ref{table2} also shows the results of the
variational calculations for \eq{form25} with $V_{\rm D}
\rightarrow V^{\rm BFW}$ and $P$ being replaced
by $\widetilde{P}$ in \eq{red8}. Here we use 
the translationally invariant h.o. basis for the variational
functions, \cite{TRGM,RED} with the maximum value
of the total h.o. quanta $N_{\rm max}=72$.
Agreement with the Faddeev calculation is satisfactory
for the ground state. For the excited $0^+$ state,
it deteriorates since $N_{\rm max}=72$ is still insufficient.
We find a very large binding energy of 19.897 MeV
for the $3\alpha$ bound state, which is
different from the result in Refs.\,\cite{TB03} and \cite{DE03}.
This difference originates from the fact
that we have used $\widetilde{P}$ instead of $P$.
We have also calculated the overlap amplitude of the ground-state
wave function, $c_{(04)}=\langle \Psi^{[3](04)}_{L=0}|\Psi\rangle$,
where the $SU_3$ (04) shell-model wave function is expressed
in the $3\alpha$ cluster model as \cite{RED}
\begin{eqnarray}
& & \Psi^{[3](04)}_{L=0}=\left[ U_{(40)}(\bp)U_{(40)}(\bq) \right]_{(04)0}
={8 \over 15}R_{20}(p, b_1) R_{20}(q, b_2)
Y_{(00)0}(\widehat{\bp},\widehat{\bq}) \nonumber \\
& & - {4 \over 3\sqrt{5}}R_{12}(p, b_1) R_{12}(q, b_2)
Y_{(22)0}(\widehat{\bp},\widehat{\bq})
+ {3 \over 5}R_{04}(p, b_1) R_{04}(q, b_2)
Y_{(44)0}(\widehat{\bp},\widehat{\bq})\ . \nonumber \\
\label{res7}
\end{eqnarray}
Here, $U_{(40)}(\bp)$ and $U_{(40)}(\bq)$ are
the single-particle $SU_3$ states
with the $SU_3$ coupling $(40) \times (40) \rightarrow (04)$,
$R_{n\ell}(x,\nu)$ is the radial part of the h.o. wave function,
and $Y_{(\lambda \ell)L}(\widehat{\bp},\widehat{\bq})$ is
the coupled angular-momentum function. The width parameters
of the h.o. radial wave functions in the momentum representation are
given by $b_1=1/4\gamma$ and $b_2=3/16\gamma$ with
$\gamma=\mu\nu=2\nu$.
For the present calculation, we have used the h.o. width
parameter $\nu=0.28125~\hbox{fm}^{-2}$ in the
configuration space.\footnote{This $\nu$ value corresponds
to a rather compact $\alpha$ cluster
with the rms radius $r_\alpha=(3/4\sqrt{\nu})
=1.414~\hbox{fm}$.}
Since the binding energy is very large, the $c_{(04)}$ value is
very close to 1, and is about 0.96.
The squared norm of the redundant components defined
by $\langle f|f \rangle$ with $|f\rangle=\langle u_B|\Psi \rangle
=\langle u_B|(1+S)|\psi \rangle$ are also shown in the last column
in Table \ref{table2}.

\begin{table}[ht]
\caption{The same as Table \protect\ref{table4},
but for the 3$\alpha$ OCM, using the Kukulin's method
of orthogonalizing pseudo-potentials (namely, \protect\eq{form42})
for the h.o. Pauli-forbidden states $|u\rangle$.
The $\lambda$ parameter, $\lambda=10^7$ MeV, is used
to eliminate the h.o. Pauli-forbidden components.
The parenthesized numbers indicate the values for the modification
of the last two digits when $\lambda=10^5$ MeV is used.
}
\label{table5}
\begin{center}
\renewcommand{\arraystretch}{1.2}
\setlength{\tabcolsep}{4mm}
\begin{tabular}{cccc}
\hline
$N_{\rm max}$ & $E_{2\alpha}$ (MeV) & $\varepsilon_{2\alpha}$ (MeV)
    & $E_{3\alpha}$ (MeV) \\
\hline
4  &   9.81970 (56)  & $6.25 \times 10^6$ ($6.25 \times 10^4$)
   & $1.87 \times 10^7$ ($1.87 \times 10^5$) \\
6  &   6.23729 (08)  & $3.02 \times 10^7$ ($3.02 \times 10^4$)
   & $9.07 \times 10^6$ ($9.06 \times 10^4$) \\
8  &   2.34421 (01)  & 14.0067 (69) & $-22.6177$ (88) \\
10 &   0.85836 (16)  & 14.3470 (77) & $-22.6745$ (56) \\
12 & $-0.11812$ (32) & 12.9759 (68) & $-25.2357$ (70) \\
14 & $-0.67019$ (37) & 12.7748 (58) & $-25.2768$ (80) \\
16 & $-1.03181$ (98) & 12.4909 (19) & $-25.5086$ (98) \\
18 & $-1.26800$ (17) & 12.3918 (29) & $-25.5244$ (56) \\
20 & $-1.43122$ (38) & 12.3215 (26) & $-25.5528$ (40) \\
30 & $-1.77280$ (95) & 12.2608 (19) & $-25.5649$ (61) \\
40 & $-1.86319$ (33) & 12.2584 (95) & $-25.5652$ (64) \\
50 & $-1.89424$ (38) & 12.2583 (94) & $-25.5653$ (65) \\
60 & $-1.90664$ (78) & 12.2583 (94) & $-25.5653$ (65) \\
\hline
\end{tabular}
\end{center}
\end{table}

Table \ref{table3} shows results of the $3\alpha$ Faddeev
calculations when the Coulomb force is switched off.
We find that the Coulomb contribution in the present $3\alpha$
ground state with the BFW potential is 7.85 MeV, which
implies that our $3\alpha$ ground state is rather compact
compared with the microscopic $3\alpha$ cluster model.
In the latter case, the standard value is $5 \sim 6$ MeV.
When the Coulomb force is neglected, we find that the
second $0^+$ state appears around $E_{3\alpha} \sim -6$ MeV.
In Table \ref{table3}, we also show in the upper half,
denoted by ``h.o.'', the results
when the h.o. Pauli-forbidden states are used for $|u\rangle$,
instead of the bound-state wave functions $|u_B\rangle$ of
the BFW potential.
The h.o. width parameter $\nu=0.28125~\hbox{fm}^{-2}$ is again
used for this calculation. 
We find that the $3\alpha$ ground state is less bound,
but the energy difference is only 2 MeV.
In this case, the elimination of the Pauli-forbidden states
of the $3\alpha$ system is rather easy, if we use the
translationally invariant h.o. basis in the varaiational
calculation. In Ref.\,\cite{TRGM}, we have examined the equivalence
between such a variational calculation and the present Faddeev
calculation using the $2\alpha$ RGM kernel.
Table \ref{table4} shows the results of the $3\alpha$ OCM with
pairwise orthogonality conditions (namely,
\eq{form25} with $V_{\rm D} \rightarrow V^{\rm BFW}$
and $P$ constructed from the h.o. $|u\rangle$),
using the translationally invariant h.o. basis.
The BFW potential for $\alpha \alpha$ is used
with $\hbar^2/M_N=41.7860~\hbox{MeV}\cdot \hbox{fm}^2$
and the h.o. Pauli-forbidden states
with $\nu=0.28125~\hbox{fm}^{-2}$.
The Coulomb force is switched off for simplicity.
Since the present $3\alpha$ ground state is very compact,
the convergence with respect to the increase of $N_{\rm max}$ is
very fast, and $N_{\rm max}=40$ is almost sufficient to obtain
the converged result.

\begin{table}[t]
\caption{
The same as Table \protect\ref{table4},
but for the $2\alpha$ folding potential
obtained from the Schmid-Wildermuth force
with $\nu=0.257~\hbox{fm}^{-2}$.
As for the spin-isospin dependence, an almost pure Serber force
with $X_d=2.4$ is used. For the reduced mass, the standard
value $\hbar^2/M_N=41.4711~\hbox{MeV}\cdot \hbox{fm}^2$ for
the microscopic $3\alpha$ cluster model is used.
}
\label{table6}
\begin{center}
\renewcommand{\arraystretch}{1.2}
\setlength{\tabcolsep}{4mm}
\begin{tabular}{ccccc}
\hline
$N_{\rm max}$ & $E_{2\alpha}$ (MeV) & $\varepsilon_{2\alpha}$ (MeV)
   & $E_{3\alpha}$ (MeV) & $c_{(04)}$ \\
\hline
4  &  14.3525  & $-$ & $-$ &  $-$ \\
6  &   9.1136  & $-$ & $-$ &  $-$ \\
8  &   4.9458  & 17.6002  &  $-5.8189$ &    1   \\
10 &   3.0476  & 15.2471  &  $-6.5774$ & 0.9873 \\
12 &   1.8698  & 13.1271  &  $-9.1708$ & 0.9528 \\
14 &   1.1573  & 11.8925  &  $-9.6655$ & 0.9293 \\
16 &   0.6854  & 11.0975  & $-10.1209$ & 0.9085 \\
18 &   0.3637  & 10.5721  & $-10.3130$ & 0.8946 \\
20 &   0.1348  & 10.2268  & $-10.4427$ & 0.8845 \\
30 & $-0.3903$ &  9.6254  & $-10.6095$ & 0.8661 \\
40 & $-0.5598$ &  9.5415  & $-10.6253$ & 0.8634 \\
50 & $-0.6313$ &  9.5286  & $-10.6271$ & 0.8630 \\
60 & $-0.6665$ &  9.5264  & $-10.6274$ & 0.8629 \\
72 & $-0.6882$ &  9.5259  & $-10.6274$ & 0.8629 \\
\hline
\end{tabular}
\end{center}
\end{table}

We have also examined the 3$\alpha$ OCM, using the Kukulin's method
of orthogonalizing pseudo-potentials (namely, \eq{form42})
for the h.o. Pauli-forbidden states $|u\rangle$.
The results are shown in Table \ref{table5} with respect to the cases
when the $\lambda$ parameter in \eq{form30}
is $\lambda=10^5$ MeV and $\lambda=10^7$ MeV.
For $\lambda=10^7$ MeV, we
find a complete agreement with the results
in Table \ref{table4} for $N_{\rm max} \ge 8$.
When $\lambda=10^5$ MeV is used, the energies deviate from the
values in $\lambda=10^7$ MeV in the last two digits,
as long as the model space is large enough.
This is, of course, a rather expected result,
since the h.o. basis can conveniently eliminate
the Pauli-forbidden states with some particular h.o. quanta
within the finite number of the basis states.

A conclusion derived from the above discussion is that
the deeply bound feature of the $3\alpha$ system in the BFW potential
does not change appreciably even if one uses the real bound states
for the BFW potential as the Pauli-forbidden
states $\vert u \rangle$, as long as the dominant shell-model
components are preserved in the Pauli-allowed space
by using $\widetilde{P}$.
This is because the deeply bound states have very large overlaps
with the h.o. Pauli-forbidden states.
The strongly attractive feature of the BFW potential
in the $3\alpha$ system is related to the short range nature
of this potential, in comparison with the usual folding potentials
in the OCM formalism.
As an example of the usual $3\alpha$ OCM,
we show in Table \ref{table6} the result
of the $\alpha \alpha$ folding potential using the
Schmid-Wildermuth force \cite{SW61} with $\nu=0.257~\hbox{fm}^{-2}$.
In this case, we use $\hbar^2/M_N=41.4711~\hbox{MeV}\cdot
\hbox{fm}^2$ for the microscopic $3\alpha$ model.
We find that the $3\alpha$ energy is $-10.63$ MeV and it is not
overbound, since the Coulomb energy is about 5 - 6 MeV. 
\begin{table}[b]
\caption{Decomposition of the $3\alpha$ energy $E_{3\alpha}$ into
the kinetic-energy and potential-energy contributions.
Here the h.o. $|u\rangle$ is used.
}
\label{table7}
\begin{center}
\renewcommand{\arraystretch}{1.4}
\setlength{\tabcolsep}{4mm}
\begin{tabular}{ccccc}
\hline
$V_{\alpha \alpha}$ & $\varepsilon_{2\alpha}$ (MeV)
 & $E_{3\alpha}$ (MeV) & $\langle H_0 \rangle$ (MeV)
 & $\langle V \rangle$ (MeV) \\
\hline
BFW & 12.258 & $-25.565$ & 124.680 & $-150.245$ \\
SW  &  9.526 & $-10.624$ &  78.405 &  $-89.030$ \\
\hline
\end{tabular}
\end{center}
\end{table}
\begin{table}[t]
\caption{The correlation between the $3\alpha$ ground-state
energy $E_{3\alpha}$ and the depth of the direct potential $V_0$.
For BFW the bound-state $|u_B\rangle$ is used.
}
\label{table8}
\begin{center}
\renewcommand{\arraystretch}{1.4}
\setlength{\tabcolsep}{4mm}
\begin{tabular}{cccc}
\hline
$V_{\alpha \alpha}$ & $V_0$ (MeV) & $b$ (fm) & $E_{3\alpha}$ (MeV) \\
\hline
SW~($\nu=0.257~\hbox{fm}^{-2}$) &  $-97.7$ & 2.26 & $-10.62$ \\
SW~($\nu=0.275~\hbox{fm}^{-2}$) & $-105.4$ & 2.21 & $-14.66$ \\
\hbox{BFW} (b.s. $|u_B\rangle$) & $-122.6$ & 2.13 & $-27.75$ \\
\hline
\end{tabular}
\end{center}
\end{table}

The short range nature of the BFW potential can be seen, by directly
comparing the nuclear part of the $\alpha \alpha$ potentials:
\begin{eqnarray}
& & V^{\rm BFW}=-122.6225~e^{-0.22~r^2}\ \ , \nonumber\\
& & V_{\rm D}({\rm SW})=-97.7~e^{-0.196~r^2}\ \ .
\label{res8}
\end{eqnarray}
In order to calculate the folding potential $V_{\rm D}({\rm SW})$,
we have used the formula
\begin{equation}
V^{\alpha \alpha}_{\rm D}(r)=2X_d v_0
\left({\nu \over \nu+3\kappa/4}\right)
^{3/2}~\exp \left\{ -{\kappa \nu \over \nu+3\kappa/4}~r^2
\right\}\ \ ,
\label{res9}
\end{equation}
for the effective two-nucleon
interaction $v(r)=v_0\,w\,e^{-\kappa r^2}$
with $w=W+BP_\sigma-HP_\tau-MP_\sigma P_\tau$.
We use an almost pure Serber force with
the Majorana parameter $m=0.505$, which corresponds
to $X_d=8W+4B-4H-2M=2.4$ in \eq{res9}.
The $1/e$ ranges of these potentials are $b=2.13~\hbox{fm}$ for the
BFW potential and $b=2.26~\hbox{fm}$ for the folding potential.
Since we have calculated $\varepsilon_{2\alpha}$, we can evaluate
the contributions of the kinetic-energy and the potential-energy
terms separately through the simple expression;
$\langle H_0 \rangle=2(3\varepsilon_{2\alpha}-E)$
and $\langle V \rangle=3(E-2\varepsilon_{2\alpha})$.
(See, for example, Ref.\,\cite{TRITON}.)
Table \ref{table7} clearly shows that the large $3\alpha$ energy
of the BFW potential is the result of the large cancellation
of the kinetic-energy and potential-energy contributions.
In the original $3\alpha$ OCM calculation
using the Schmid-Wildermuth force,
Horiuchi \cite{HO75} has used the
value $\nu=0.275~\hbox{fm}^{-2}$ and
the pure Serber force, $X_d=2.445$.
In this case, the direct potential becomes
\begin{equation}
V_{\rm D}({\rm SW})=-105.4~e^{-0.204~r^2}\ , 
\label{res10}
\end{equation}
with $b=2.21~\hbox{fm}$ and the converged $3\alpha$ bound-state
energy is $-14.66$ MeV.\footnote{In Ref.\,\cite{HO75} it is
reported as $-14.68$ MeV.}
We find that the $3\alpha$ energy is strongly
correlated with the depth of the direct potential $V_0$,
as shown in Table \ref{table8}.
If we extrapolate the BFW value from the above two results
of the SW force, we find $-23.68$ MeV for the BFW potential,
which is close to the calculated value $-27.75$ MeV.
It is natural that a deep potential gives stronger binding
in the $3\alpha$ system, since the effect of the potential term
is by factor 1.5 larger than in the $2\alpha$ system,
as was pointed out by Horiuchi \cite{HO75}.

\section{Summary}

In this study, we have developed the Faddeev formalism for the
three cluster systems, which ``exactly'' takes into account
the Pauli-forbidden states.
Actually, the exact Pauli-forbidden states of three-cluster systems
are defined through the eigen-value problem of the three-cluster
normalization kernel with the eigen-value zero. 
However, the pairwise orthogonality conditions to the total wave
functions developed in this paper are known
to give a good approximation
for the exact Pauli-allowed space obtained by the diagonalization
procedure of the normalization kernel.
For example, in the $3\alpha$ system composed
of the simple $(0s)^4$ harmonic-oscillator (h.o.) shell-model
wave functions, this correspondence
is completely verified by enumerating the $SU_3$ allowed states
in the translationally invariant h.o. basis. \cite{FU81}

The main result of this paper is that it is this type
of three-cluster orthogonality condition model (OCM) with
pairwise orthogonality conditions that leads to the complete
equivalence to the three-cluster Faddeev equation interacting
only by pairwise interactions. The pairwise interaction can be
two-cluster RGM kernels with the linear energy dependence,
two-cluster folding potentials of the effective two-body force,
or the deep phenomenological local potentials like the Buck,
Friedrich and Wheatley potential (BFW potential) \cite{BF77}.
In order to formulate the three-cluster Faddeev equation
with explicit elimination of the pairwise Pauli-forbidden
components, we only need to
use the modified $T$-matrix, $\widetilde{T}(\omega, \varepsilon)$,
which eliminates the off-shell singularity related to the existence
of the Pauli-forbidden states. This $T$-matrix is, in general,
energy dependent ($\varepsilon$-dependent) for the three-cluster
Faddeev equation using two-cluster RGM kernels,
and the energy dependence is self-consistently determined
by calculating the expectation value of the two-cluster
Hamiltonian with respect to the resultant Faddeev
solution.\,\cite{TRGM} On the other hand, in the three-cluster OCM,
considered in the present paper, $\widetilde{T}(\omega)$ is
energy independent. The Pauli-forbidden state in this OCM could be
the h.o. wave functions of the microscopic two-cluster system,
or the real bound states of the phenomenological local potentials
between two clusters.
However, the $3\alpha$ Pauli-allowed space should be carefully
defined, in order not to exclude the dominant $SU_3$ components
for the realistic description of the $\hbox{}^{12}\hbox{C}$ ground
state.
We find that the Kukulin's method of orthogonalizing
pseudo-potentials \cite{KU78} is completely equivalent
to the three-cluster OCM with pairwise orthogonality conditions.
The latter was first proposed by Horiuchi \cite{HO74}.
We have proven this equivalence through the equivalence between
either model and the present three-cluster Faddeev formalism
using $\widetilde{T}(\omega)$.
A nice feature of the present Faddeev formalism is that
the $T$-matrix description of the two-cluster interaction
allows us to take the limit $\lambda \rightarrow \infty$
analytically in the method of orthogonalizing pseudo-potentials,
and that the solution of the Faddeev equation
automatically guarantees the pairwise orthogonality conditions
of the total wave function, owing to the orthogonality property
of the $\widetilde{T}(\omega)$ $T$-matrix.

As an example, we have applied the present three-cluster Faddeev
formalism to the $3\alpha$ system interacting
via the BFW $\alpha \alpha$ potential \cite{BF77}.
The Pauli-forbidden states are assumed to be
the real bound states of the potential.
We have found that this potential yields
the $3\alpha$ ground-state energy $-19.897$ MeV,
which is a different result from Refs.\,\cite{TB03} and \cite{DE03},
in which the orthogonality to the $2\alpha$ bound-state
solutions $|u_B\rangle$ is very strictly demanded.
This feature of the large overbinding does not change even when
we use the h.o. Pauli-forbidden states with a reasonable width
parameter of the $\alpha$-clusters. This feature is traced back
to the deep and short-range nature
of the BFW $\alpha \alpha$ potential,
as compared with the folding potentials usually used
in the microscopic $3\alpha$ cluster model.
The validity of the $3\alpha$ boson model described
by the BFW $\alpha \alpha$ potential is examined by calculating
other physical observables, like the rms radius
of $\hbox{}^{12}\hbox{C}$, using the obtained $3\alpha$ ground-state
wave function. 

\begin{acknowledge}
This work was supported by Grants-in-Aid for Scientific
Research (C) from the Japan Society for the Promotion
of Science (JSPS) (Nos. 15540270, 15540284).
Y. Fujiwara wishes to thank the FNRS foundation
of Belgium for making his visit to the Free University of Brussels
possible during the summer, 2002.
The discussion with Professor D. Baye is greatly appreciated.
\end{acknowledge}

\appendix
\section*{A simple formula for \mib{T}-matrix}
\setcounter{section}{1}

In this Appendix, we will show a general formula of the $T$-matrix
for the sum of two potentials. From a simple calculation,
the $T$-matrix for $V=V_1+V_2$ is expressed as $T=T_1+T_2$ with
\begin{equation}
T_1=t_1+t_1 G_0 T_2 \ ,\qquad
T_2=t_2+t_2 G_0 T_1 \ ,
\label{a1}
\end{equation}
where $t_1$ and $t_2$ are the $T$-matrices
of $V_1$ and $V_2$, respectively:
\begin{equation}
t_1=V_1+V_1 G_0 t_1 \ ,\qquad
t_2=V_2+V_2 G_0 t_2 \ .
\label{a2}
\end{equation}
If we further assume $V_2=\lambda\,|u\rangle \langle u|$,
we find
\begin{equation}
t_2=|u\rangle {1 \over \lambda^{-1}-\langle u|G_0|u \rangle}
\langle u| \ .
\label{a3}
\end{equation}
Thus, the solution of~$T=V+V G_0 T$~for
~$V=V_1+\lambda |u\rangle \langle u|$~is given by
\begin{eqnarray}
& & T=t_1+(1+t_1 G_0)|u\rangle
{1 \over \lambda^{-1}-\langle u|G_0+G_0 t_1 G_0|u \rangle}
\langle u|(1+G_0 t_1) \ . 
\label{a4}
\end{eqnarray}
If we further move to $\lambda \rightarrow \infty$, we find
\begin{eqnarray}
& & T=t_1-(1+t_1 G_0)|u\rangle
{1 \over \langle u|G_0+G_0 t_1 G_0|u \rangle}
\langle u|(1+G_0 t_1) \ ,
\label{a5}
\end{eqnarray}
with $t_1=V_1+V_1 G_0 t_1$.
This expression leads to our basic relationship
\begin{equation}
\langle u|\left[1+G_0 T\right]=0
\qquad \hbox{and} \qquad \left[1+T G_0\right]|u \rangle=0 \ .
\label{a6}
\end{equation}

As an example, let us use this formula in the method of
orthogonalizing pseudo-potentials for $V_{\rm RGM}(\varepsilon)$.
We set
\begin{eqnarray}
V_1 \rightarrow V_{\rm RGM}(\varepsilon) \qquad \hbox{and} \qquad
t_1 \rightarrow T(\omega, \varepsilon) \ \ .
\label{a7}
\end{eqnarray}
If we use the relationship between $T(\omega, \varepsilon)$
and $\widetilde{T}(\omega, \varepsilon)$,
given in  Eqs.\,(2.8), (2.9),
and (2.10) of Ref.\,\cite{TRGM}, we easily obtain 
\begin{eqnarray}
T & = & T(\omega, \varepsilon)-\left[1+T(\omega, \varepsilon)
G_0(\omega) \right]|u\rangle
{1 \over \langle u|G_0(\omega)+G_0(\omega)T(\omega, \varepsilon)
G_0(\omega)|u \rangle} \nonumber \\
& & \times \langle u|\left[1+G_0(\omega)T(\omega, \varepsilon)
\right] \nonumber \\
& = & T(\omega, \varepsilon)-(\omega-H_0)|u\rangle
{1 \over \omega-\varepsilon}\langle u|(\omega-H_0)
=\widetilde{T}(\omega, \varepsilon) \ .
\label{a8}
\end{eqnarray}
Namely, our RGM $T$-matrix $\widetilde{T}(\omega, \varepsilon)$
is obtained from $V_{\rm RGM}(\varepsilon)$ by the method
of orthogonalizing pseudo-potentials.

\end{document}